%% file: main.tex
\documentclass[10pt,sigconf]{acmart}

\usepackage{balance}

\usepackage{booktabs} %
\usepackage{cleveref}
\usepackage{url}

\graphicspath{{./figures/}}
\usepackage{graphicx}
\usepackage{subcaption}

\usepackage{xspace}
\newcommand*{\eg}{e.g.,\@\xspace}
\newcommand*{\ie}{i.e.,\@\xspace}
\newcommand*{\cf}{cf.\@\xspace}

\newcommand{\ak}[1]{}
\newcommand{\tim}[1]{}

\def\ParHead{\vspace*{2mm}\noindent\bf}

\AtBeginDocument{%
  \providecommand\BibTeX{{%
    \normalfont B\kern-0.5em{\scshape i\kern-0.25em b}\kern-0.8em\TeX}}}

\copyrightyear{2020} 
\acmYear{2020} 
\setcopyright{acmlicensed}\acmConference[aiDM'20]{Third Workshop in Exploiting AI Techniques for Data Management }{June 14--19, 2020}{Portland, OR, USA}
\acmBooktitle{Third Workshop in Exploiting AI Techniques for Data Management (aiDM'20), June 14--19, 2020, Portland, OR, USA}
\acmPrice{15.00}
\acmDOI{10.1145/3401071.3401659}
\acmISBN{978-1-4503-8029-4/20/06}

\begin{document}

\title{RadixSpline: A Single-Pass Learned Index}

\author{Andreas Kipf$^{\,\star}$ \ \ Ryan Marcus$^{\,\star\dagger}$ \ \ Alexander van Renen \ \ Mihail Stoian}
\author{Alfons Kemper \ \ Tim Kraska$^{\,\star}$ \ \ Thomas Neumann}
\affiliation{TUM \ \ \ \ MIT CSAIL$^{\,\star}$ \ \ \ \ Intel Labs$^{\,\dagger}$}
\affiliation{\small \{renen, stoian, kemper, neumann\}@in.tum.de \ \ \ \ \{kipf, ryanmarcus, kraska\}@mit.edu}

\thanks{Andreas Kipf, Ryan Marcus, and Alexander van Renen contributed equally}

\renewcommand{\shortauthors}{Kipf et al.}

\input{abstract}

\maketitle

\input{introduction}

\input{radixspline}

\input{evaluation}
\input{conclusions}

\bibliographystyle{abbrv}
\bibliography{ryan-cites-long}

\end{document}

%% file: abstract.tex
\begin{abstract}
Recent research has shown that learned models can outperform state-of-the-art index structures in size and lookup performance.
While this is a very promising result, existing learned structures are often cumbersome to implement and are slow to build.
In fact, most approaches that we are aware of require multiple training passes over the data.

We introduce RadixSpline (RS), a learned index that can be built in a single pass over the data and is competitive with state-of-the-art learned index models, like RMI, in size and lookup performance.
We evaluate RS using the SOSD benchmark and show that it achieves competitive results on all datasets, despite the fact that it only has two parameters.
\end{abstract}

%% file: introduction.tex
\section{Introduction}

\begin{figure}
    \centering
    \includegraphics[width=0.49 \textwidth]{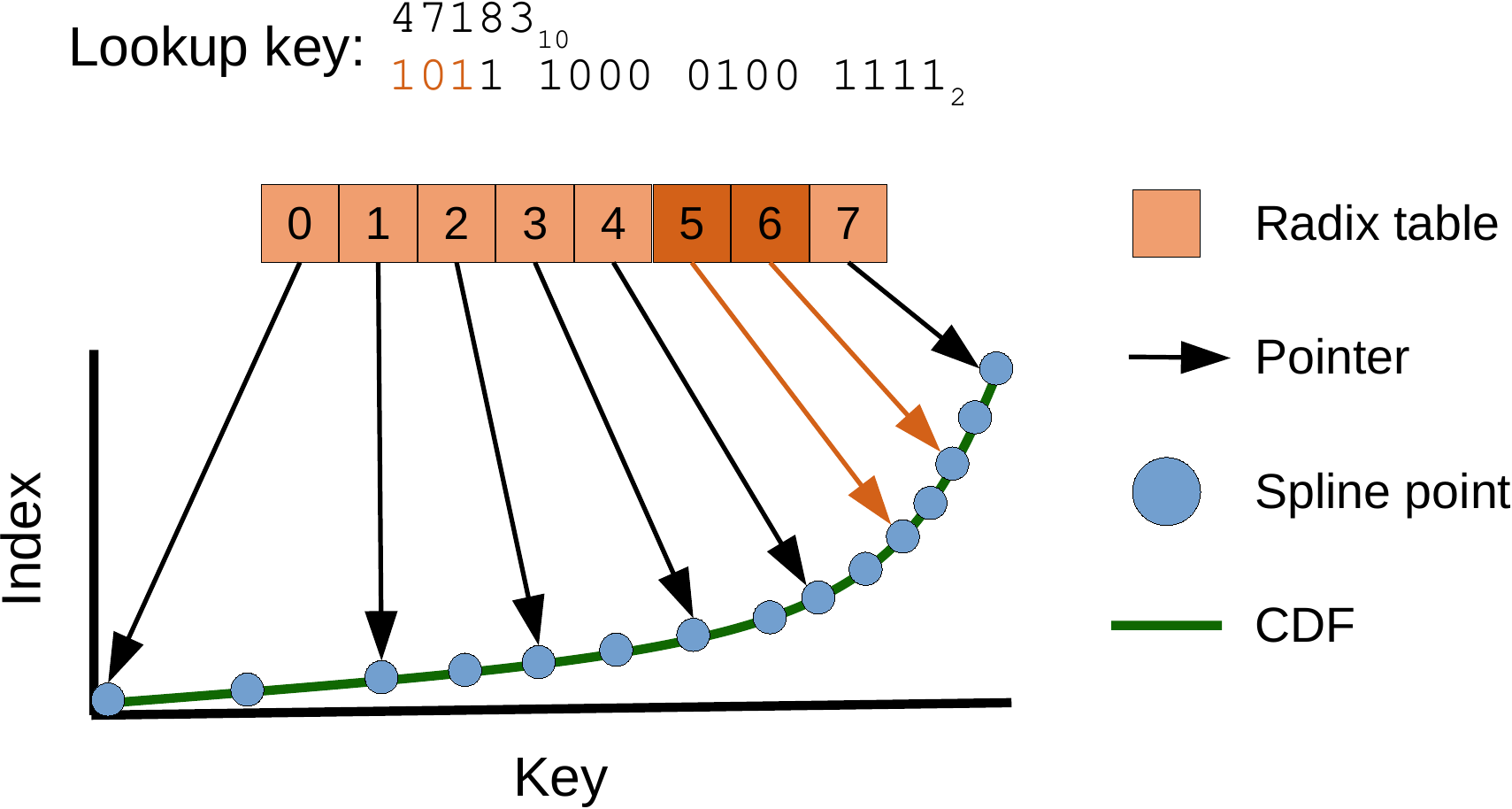}
    \caption{A radix spline index and example lookup process. The $r$ (here, 3) most significant bits $b$ of the lookup key are used as an index into the radix table. Then, a binary search is performed on the spline points between the $b$th pointer and the $b+1$th pointer.}
    \label{fig:rs}
\end{figure}

In ~\cite{ml_index}, Kraska et al. proposed learned index structures, a new type of index for sorted data which use learned models to predict the position of a lookup key. These learned index structures can be realized via supervised learning techniques, using the cumulative distribution function (CDF) of the underlying data for training. More recently, the SOSD benchmark~\cite{sosd} demonstrated that learned index structures (which can be viewed as CDF approximators) can compete favorably with state-of-the-art index structures~\cite{art,start,hot,fast,surf,ibtree} in terms of size and lookup performance.

However, some learned index structures, such as RMIs~\cite{ml_index}, do not support inserts and cannot be constructed in a single pass over the data, which severely limits their applications.
The recent learned index proposals ALEX~\cite{alex} and PGM~\cite{pgm-index} add support for inserts.
In this work, we argue that there are applications where indexes do not need to support individual updates and where it is sufficient to be able to build them efficiently.
The most prominent example are LSM-trees~\cite{lsm}.

In an LSM-tree, data is stored in several files, each of which is sorted by a key column. Each file generally stores additional metadata, such as a Bloom filter or an index. This metadata can be used at query time to either exclude a file from a lookup (via the Bloom filters or lower/upper bound) or to quickly locate the relevant tuples in a file (via the index). These files are organized into multiple levels, with higher levels containing exponentially more files than lower levels. New data is inserted into files in lower levels, which are periodically merged with files in higher levels. Unfamiliar readers may wish to see~\cite{lsm_survey} for an in-depth survey of LSM-trees, but for this work one only needs to note that this merge process between two files is the perfect time to re-build a learned index.
The merge produces data in sorted order, which can be passed through a \emph{single-pass} training algorithm before it is written back to disk.
Since the merge operation is expensive on its own and is usually done asynchronously, training such a one-pass learned index could only incur a negligible constant overhead.
However, existing learned indexes do not allow for an efficient build.

In this work, we introduce RadixSpline (RS), a learned index that can be built in a single pass over sorted data.
Notably, the proposals of FITing-Tree~\cite{fiting_tree} and PGM~\cite{pgm-index} also support single-pass builds.
However, for both of these indexes the amount of work per element is logarithmic in the number of levels (similar to inserts in a BTree).
With RS, we propose the first single-pass learned index with a \emph{constant} amount of work per new element.

Being an ordered index, RS supports both equality and range predicates (\eg lower bound lookups).
RS is built in two steps.
First, a linear spline is fit to the CDF of the data that guarantees a certain error bound. This results in a set of spline points which can be significantly smaller than the underlying data.
Second, we build a radix table (a flat radix structure) that serves as an approximate index into the spline points.
Similar to the \texttt{Node256} in the Adaptive Radix Tree (ART)~\cite{art}, we extract a certain radix prefix (\eg the first $20$~bits, neglecting common prefix bits shared by all keys) and use those as an offset into the radix table.
Both steps can be performed in a single pass over the sorted data.

RS is not only efficient to build, but also competitive with state-of-the-art RMI models in size and lookup performance.
Index size is an especially important factor for LSM-tree applications because indexes are kept in main memory (whereas each large sorted file is stored on disk).
Furthermore, RS's implementation only consists of roughly one hundred lines of C++ code and does not have any external dependencies.
Finally, RS only takes two hyper parameters (spline error and radix table size). Both have an intuitive and reliable impact on size and lookup latency. As a result, tuning RS is easier than tuning more complex learned indexes structures with many hyperparameters~\cite{cdfshop}.
One caveat is that RS can be impacted by heavy skew, rendering the radix table largely ineffective.
In such cases, one could fall back to a tree-structured radix table or handle outliers separately.
However, we are yet to encounter such extreme skew in real-world data.
In summary, we believe that RS is a practical learned index structure with potentially high impact in write-once/read-many settings such as LSM-trees.

More broadly, this work follows recent trends in integrating machine learning components into systems~\cite{pillars}, especially database systems~\cite{neo, deep_card_est2, bao, rejoin, cidr_dlqo, naru, plan_loss, skinnerdb, qo_state_rep, deep_sketch, learn_cost, local_card_est, sql_embed, ml_tuning}.

%% file: radixspline.tex
\section{RadixSpline}
\label{sec:rs}

A RadixSpline (RS) index is designed to map a \emph{lookup key} to an \emph{index} (the position of the key in the underlying data).
Like an RMI~\cite{ml_index}, radix spline indexes require the underlying data to be sorted on the lookup key in a flat array.

An RS index consists of two components: a set of spline points and a radix table. The set of spline points is a subset of the keys, selected so that spline interpolation for any lookup key will result in a predicted lookup location within a preset error bound. For example, if the preset error bound is 32, then the location of any lookup key can be no more than 32 positions away from the location predicted by the RS index. The radix table helps to quickly locate the correct spline points for a given lookup key. Intuitively, the radix table limits the range of possible spline points to search over for every possible $b$-length prefix of a lookup key.

At lookup time, the radix table is used to determine a range of spline points to examine. These spline points are searched until the two spline points surrounding the key are found. Then, linear interpolation is used to predict the location (index) of the lookup key in the underlying data. Because the spline interpolation is error-bounded, only a (small) range of the underlying data needs to be searched.

In contrast to other learned indexes~\cite{ml_index,alex}, RS can be built in a single pass over the sorted data.
While the use of splines and a bottom-up approach has been explored before in FITing-Tree~\cite{fiting_tree} and others~\cite{pgm-index,lis_func_interp}, in this work, we combine the ideas from~\cite{fiting_tree} with radix trees, making it highly competitive with top-down built indexes~\cite{ml_index}.

\subsection{Construction}
RS indexes, like PGM indexes~\cite{pgm-index} or FITing-Trees~\cite{fiting_tree}, are built ``bottom-up''. First, we construct an error-bounded spline on top of the the underlying data. Then, the selected spline points themselves are indexed in a radix table.

\begin{figure}
    \centering
    \includegraphics[width=0.4 \textwidth]{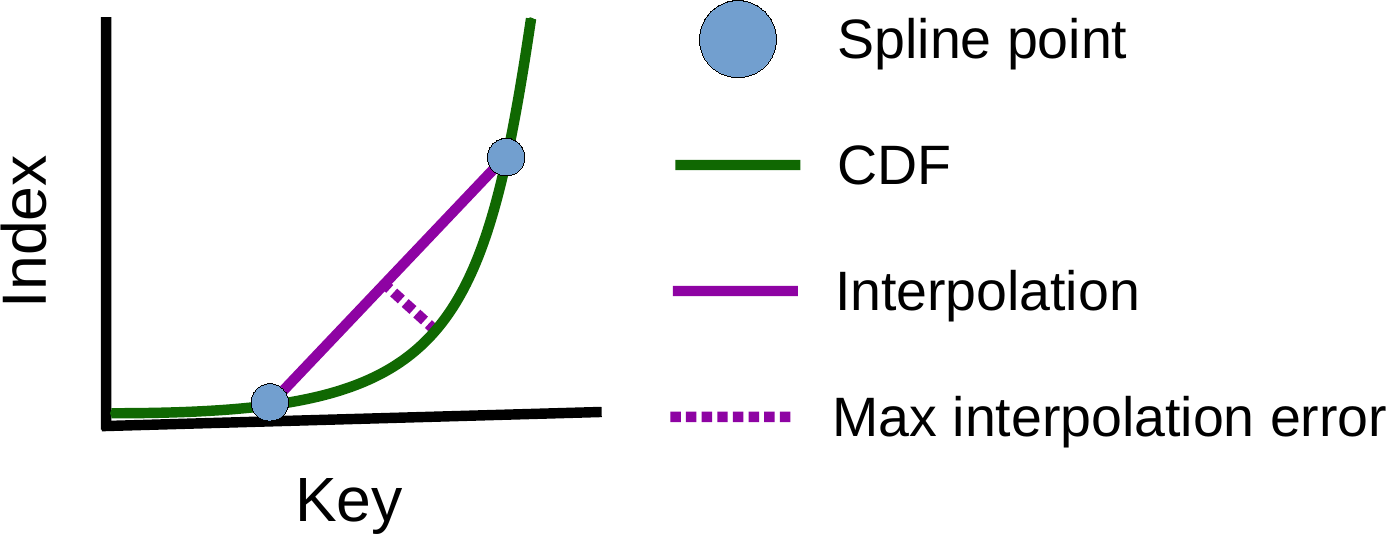}
    \caption{A single spline segment. We select enough spline points so that the maximum interpolation error (dashed line) stays within a specified bound.}
    \label{fig:spline_error}
\end{figure}
{\ParHead Build Spline.}
As observed in~\cite{ml_index}, all index structures can be thought of as models that map lookup keys to positions. Let the dataset to index $D$ be an indexed set of tuples, with $D_i = (k_i, p_i)$ where $D_i$ represents the $i$th datapoint, $k_i$ represents the key of the $i$th datapoint, and $p_i$ represents the position (offset) of the $i$th datapoint. A radix spline index first builds a spline model $S$, such that $S(k_i) = p_i \pm e$, where $e$ is a specified constant. In other words, the spline model $S$ \emph{always} predicts the correct location of the data within a constant error of $e$.

This error-bounded model is realized via spline interpolation (we use \texttt{GreedySplineCorridor}~\cite{spline}). The parameters of the model $S$, $Knots(S)$, are a set of spline points, or knots, which are a representative set of datapoints (see Figure~\ref{fig:rs}). These data points are chosen such that, for any lookup key $x$, linearly interpolating between the two closest spline points in $Knots(S)$ will produce an estimate with error no larger than $e$ (\cf Figure~\ref{fig:spline_error}). Formally, to evaluate $S(x)$, letting $(k_{\mathit{left}}, p_{\mathit{left}}) \in Knots(S)$ be the knot with the greatest key such that $k_{\mathit{left}} \leq x$ and letting $(k_{\mathit{right}}, p_{\mathit{right}}) \in Knots(S)$ be the knot with the smallest key such that $k_{\mathit{right}} > x$, we compute

\begin{equation*}
S(x) = p_{\mathit{left}} + (x - k_{\mathit{left}}) \times \frac{p_{\mathit{right}} - p_{\mathit{left}}}{k_{\mathit{right}} - k_{\mathit{left}}}.
\end{equation*}

For more details on the error-bounded spline algorithm, we refer the reader to~\cite{spline}.

\begin{figure*}[ht!]
\centering
{\includegraphics[width=0.33\linewidth]{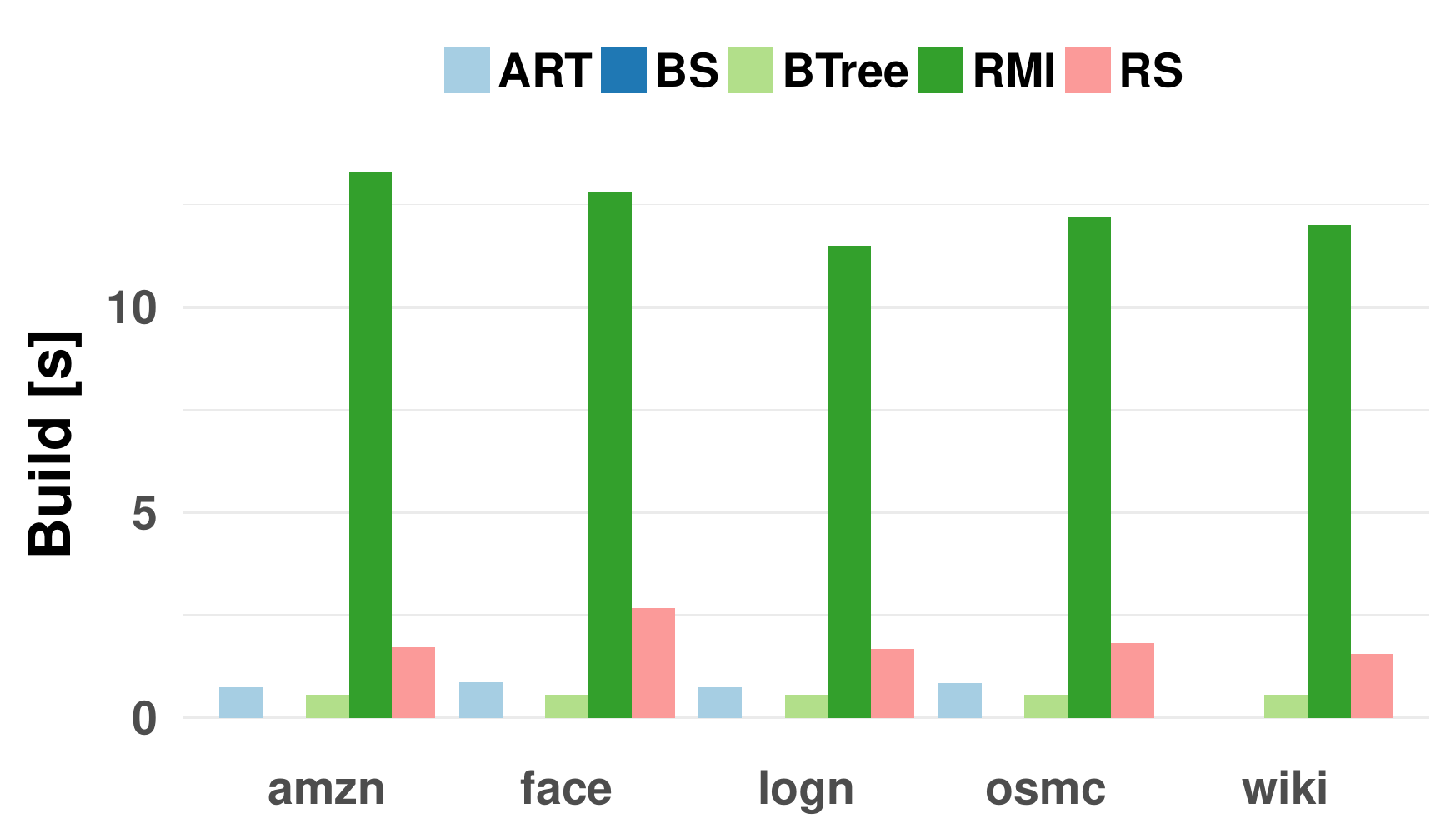}}
{\includegraphics[width=0.33\linewidth]{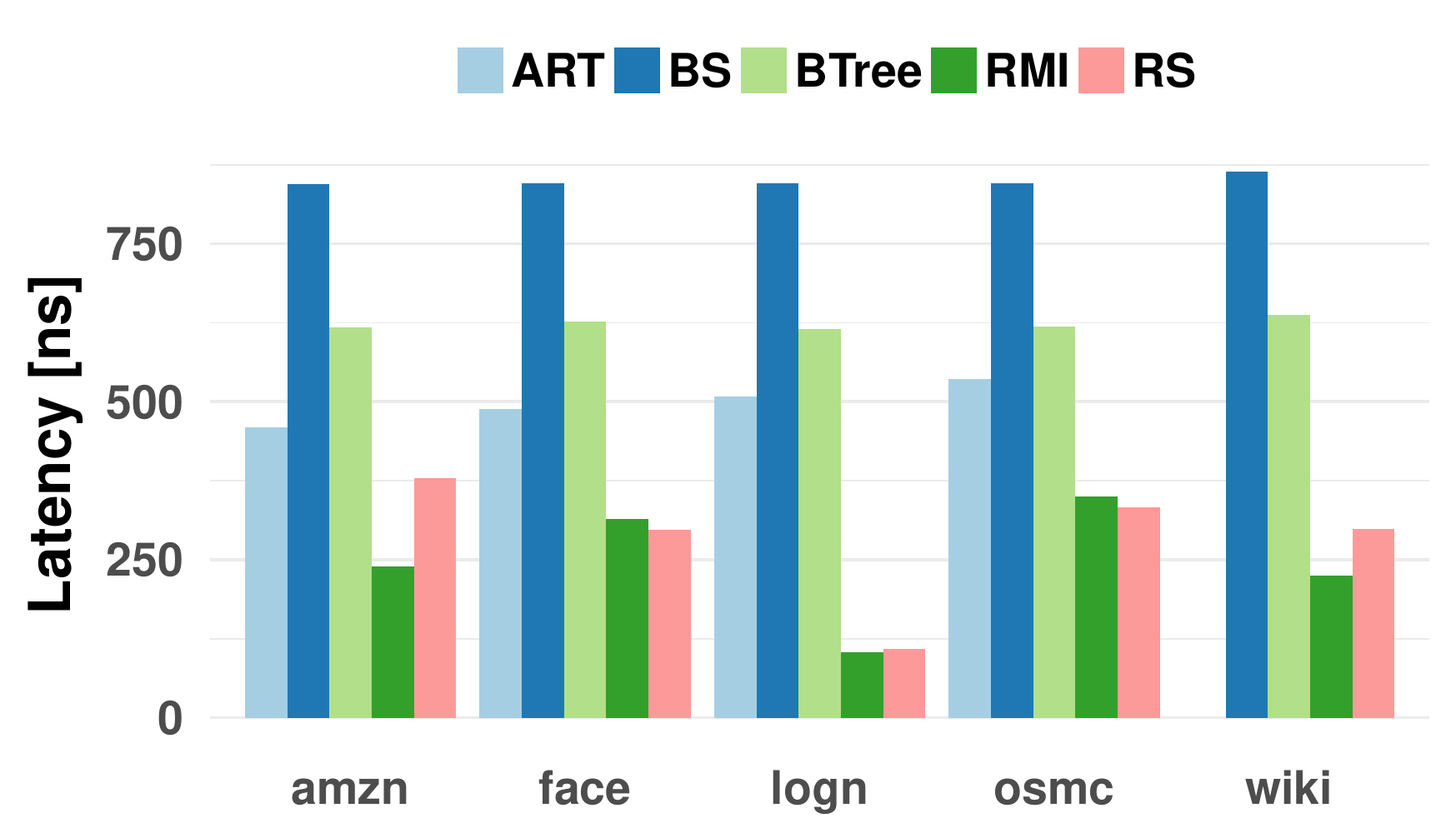}}
{\includegraphics[width=0.33\linewidth]{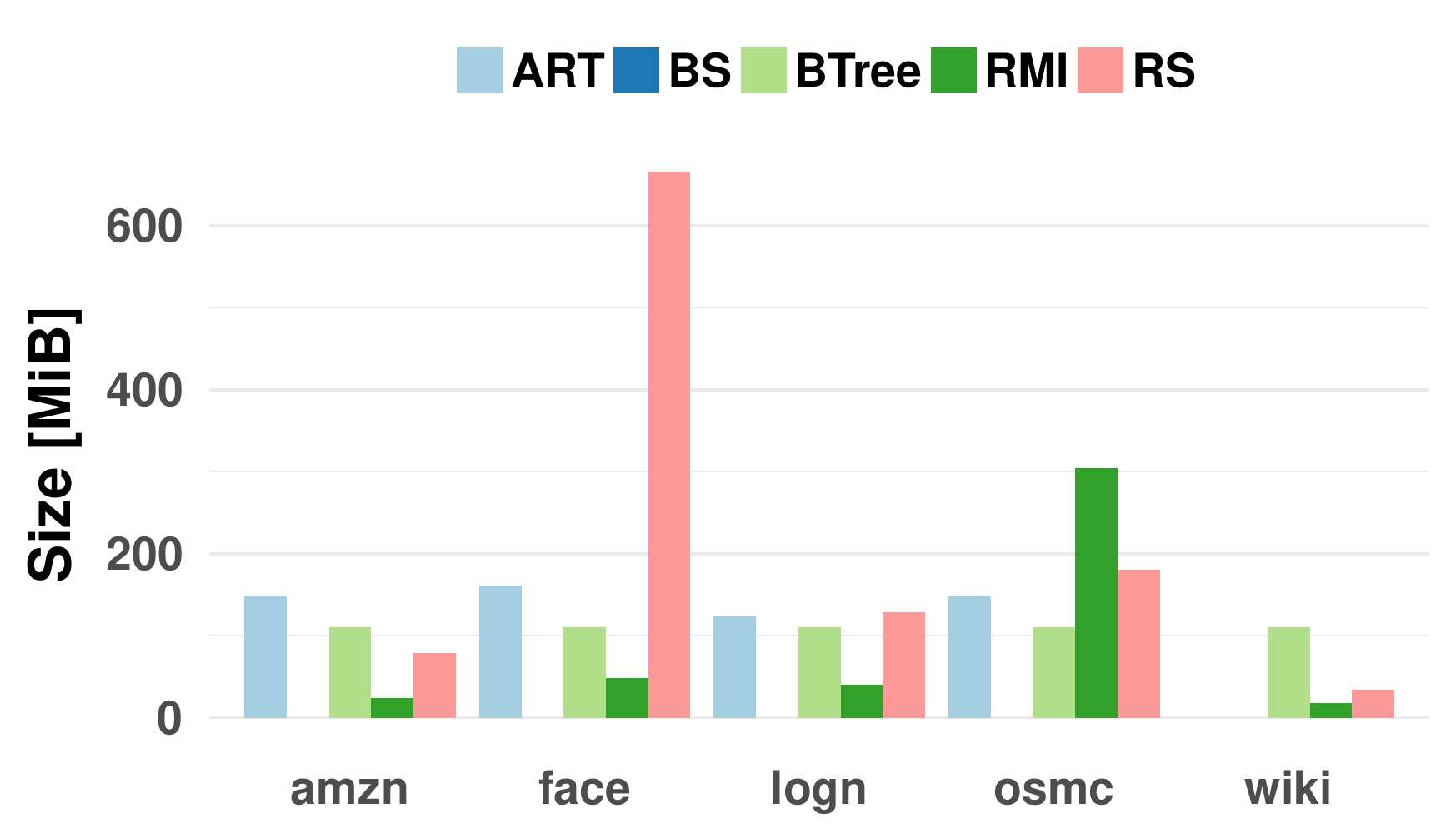}}
\vspace{-6mm}
\caption{Build time (left), lookup latency (middle), and size (right) for lookup-optimized index configurations.}
\vspace{-2mm}
\label{fig:results}
\end{figure*}

{\ParHead Build Radix Table.}
Next, we build a radix table on top of the selected spline points to quickly find the two spline points surrounding the lookup key.
The radix table is a flat \texttt{uint32\_t} array that maps fixed-length key prefixes (``radix bits'') to the first spline point with that prefix.
The key prefixes are the offsets into the radix table while the spline points are represented as \texttt{uint32\_t} values stored in the radix table (\cf pointers in Figure~\ref{fig:rs}).

The radix table takes the number of radix bits $r$ as a parameter.
For example, for $r = 18$ we allocate an array with $2^{18}$ many entries (1\,MiB in size).
A larger $r$ grows the size of the table exponentially ($2^r$) but may also increase its precision.
That is, we may need to search a more narrow range of spline points to find the two spline points surrounding the lookup key.
In Section~\ref{sec:evaluation}, we show the impact of this parameter on size and lookup performance.

The build process itself is very straightforward and extremely fast: we first allocate an array of the appropriate size ($2^{r}$ many entries), then we go through all spline points and whenever we encounter a new $r$-bit prefix $b$, we insert the offset of the spline point (a \texttt{uint32\_t} value) into the slot at offset $b$ in the radix table.
Since the spline points are ordered, the radix table is filled in consecutive order from left to right.
As an optimization, we eliminate common prefix bits shared by all keys when building the radix table.

{\ParHead Single Pass.}
Building the CDF, the spline, and the radix table can all be performed on-the-fly, in a single pass over the sorted datapoints.
When encountering a new CDF point (\ie when the key changes), we pass that point to the spline construction algorithm~\cite{spline}.
Filling the pre-allocated radix table within the same pass is also straightforward: whenever we encounter a new $r$-bit prefix in a selected spline point, we make a new entry to the table.

{\ParHead Lookups.}
Using the example in Figure~\ref{fig:rs}, the lookup logic is as follows:
We first extract an $r$-bit prefix $b$ of the lookup key ({\color{orange} 101} in this case).
Then, we use the extracted bits $b$ to make an offset access into the radix table retrieving the two pointers stored at positions $b$ and $b+1$ (here, positions 5 and 6).
These pointers (marked in orange) define a narrowed search range on the spline points.
Next, we search this range for the two spline points surrounding the lookup key using binary search.
Subsequently, we perform a linear interpolation between these two spline points to obtain an estimated position $p$ of the key.
Finally, we perform a binary search within the error bounds ($p \pm e$) to find the first occurrence of the key.

%% file: evaluation.tex
\section{Evaluation}
\label{sec:evaluation}

We evaluate RadixSpline (RS) using the SOSD benchmark~\cite{sosd} on a \texttt{c5.4xlarge} AWS machine.
We use six 64-bit datasets, each of them containing 200\,M key/value pairs (3.2\,GiB) in size: \texttt{amzn} (book popularity data), \texttt{face} (Facebook user IDs), \texttt{logn} (synthetic lognormal distributed data), \texttt{osmc} (composite cell IDs from Open Street Map) and \texttt{wiki} (timestamps of Wikipedia edits).
For details on these datasets, see~\cite{sosd}.

Like~\cite{sosd}, we build indexes on top of sorted arrays. An index takes in a key and produces a \emph{search range} in the underlying data. This range must contain the lookup key if the lookup key exists, and must otherwise contain the first key not larger than the lookup key (lower bound search). Then, binary search is used to locate the exact key within the search range. Indexes are evaluated based on their end-to-end performance: the time to produce a search range plus the time to execute the binary search. We perform 10M lookups (1 thread) on a given dataset and report the average lookup latency. Lookup keys are uniformly chosen from the keys.

We compare RS against three traditional, non-learned approaches: ART~\cite{art}, STX B+-tree (BTree)~\cite{url-stxbtree}, and binary search (BS). For ART and BTree, we use a stride of 32 (meaning that every 32nd key is inserted into the index -- this provides better space and performance compared to indexing each key). We also compare against the public implementation~\cite{cdfshop} of the recursive model index (RMI)~\cite{ml_index}, a learned approach that is built ``top-down'' (\ie starting with a loose fit and then progressively learning finer-grained models) and internally uses a range of models (\eg linear, cubic, or even BTrees). Hash-based methods are excluded because hash-based methods do not support lower bound searches. Since ART does not support duplicate keys, it does not have results for \texttt{wiki} (the only dataset with duplicates).

{\ParHead Build Times.}
(Figure~\ref{fig:results}, left).
Due to its single-pass build process, RS is almost as efficient to build as ART or BTree and is significantly faster than RMI, which performs multiple training passes over the sorted datapoints.

\begin{figure*}[ht!]
\centering
{\includegraphics[width=0.33\linewidth]{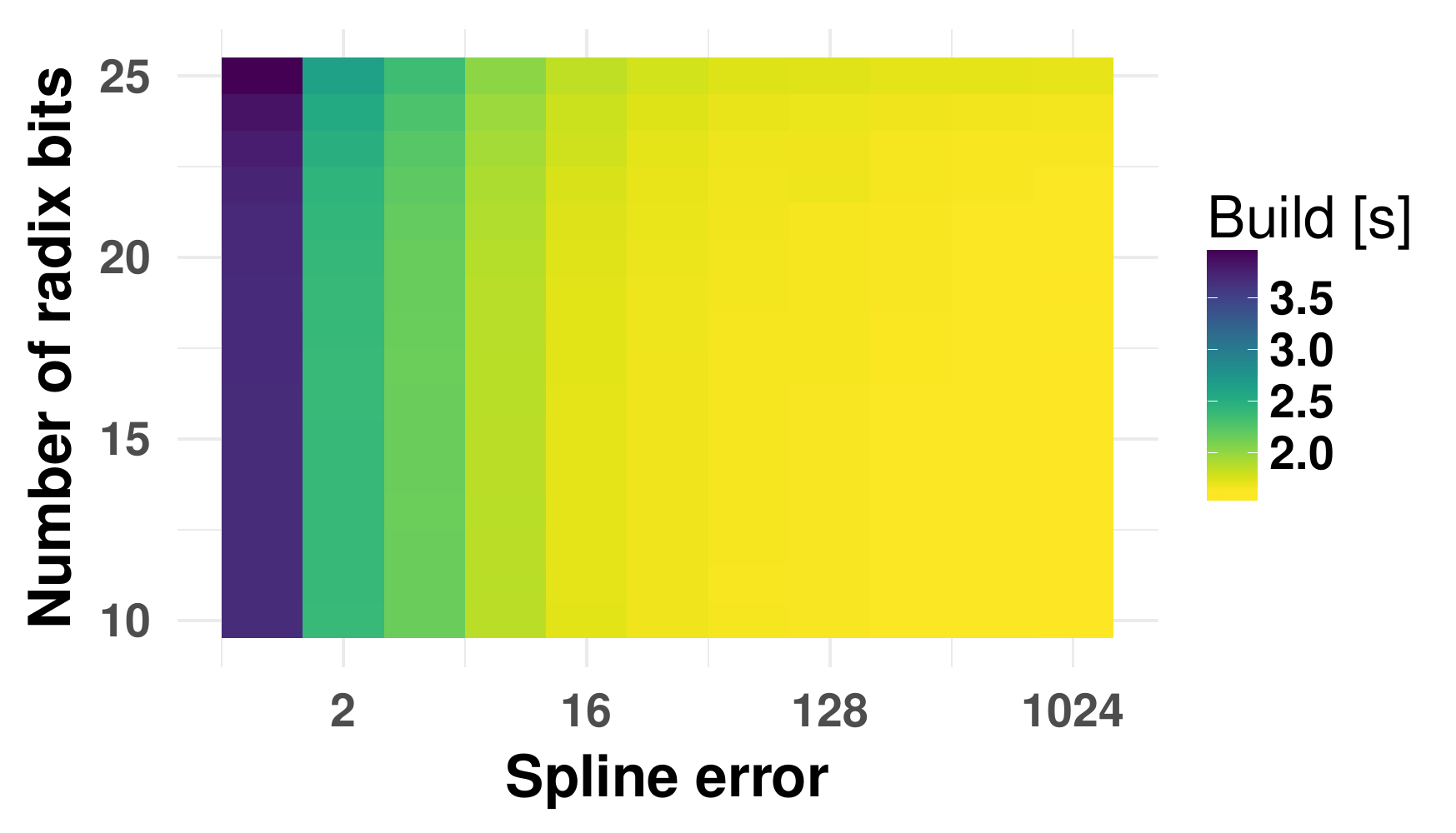}}
{\includegraphics[width=0.33\linewidth]{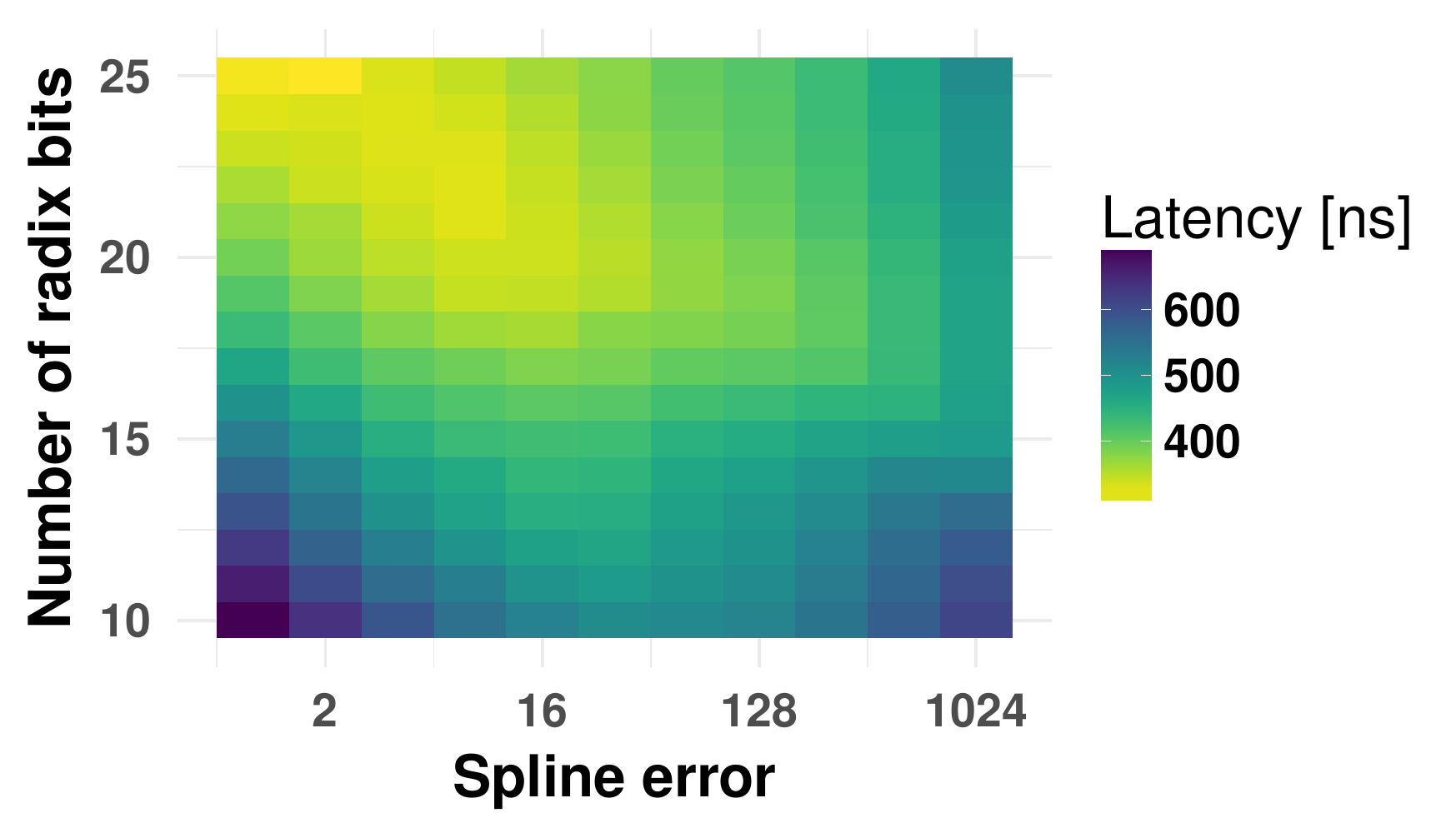}}
{\includegraphics[width=0.33\linewidth]{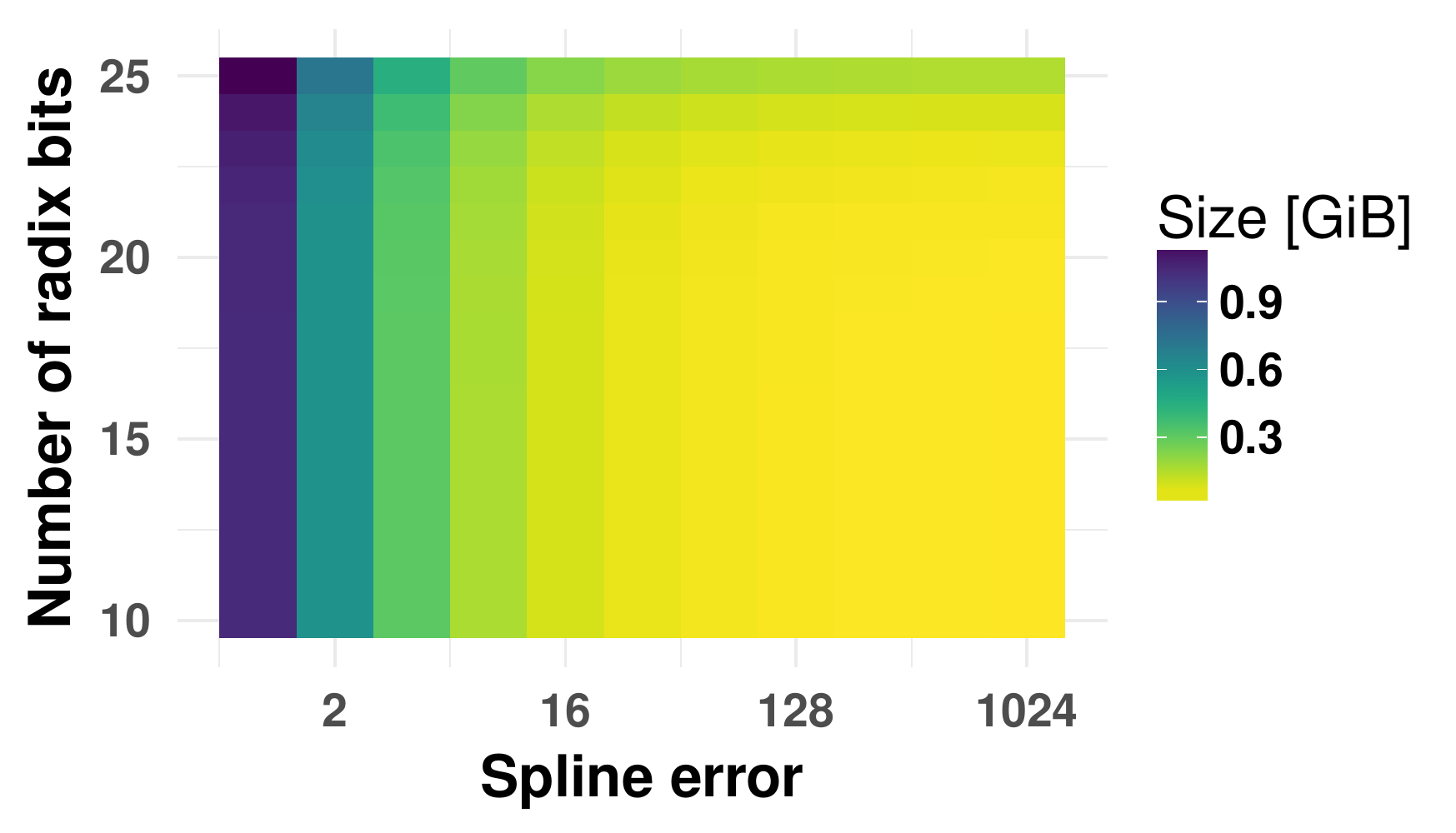}}
\caption{Build time (left), lookup latency (middle), and index size (right) for different RadixSpline configurations.}
\label{fig:heat}
\end{figure*}

{\ParHead Lookup Latency.}
(Figure~\ref{fig:results}, middle).
Binary search (BS) takes around 850ns per lookup across all datasets.
BTree improves upon BS by using the cache more efficiently, and requires only around 600ns. Like BS, it is largely independent of the data distribution.
Both learned approaches, RMI and RS, are significantly faster than the traditional indexes but also more affected by the data distribution.
Note that we have tuned both approaches for minimum lookup latency.

{\ParHead Index Size.}
(Figure~\ref{fig:results}, right).
Except for BS and a few outliers, all indexes consume around 100\,MiB which corresponds to 6.6\% of the uncompressed key size (200\,M 64-bit keys).
For the \texttt{face} dataset, RS uncharacteristically requires more than 600\,MiB (39\%) to achieve its best performance.
Next, we investigate different RS configurations for this dataset.

{\ParHead Configuration Space.}
The best RS configuration (Figure~\ref{fig:results}) for \texttt{face} uses a lot of memory ($\approx 650$\,MiB). However, we can easily trade lookup performance for memory as shown in Figure~\ref{fig:heat}:
Both build time (left) and size (right) mostly depend on the spline error.
Starting with an error of 16, RS can be built within 2s for this dataset and requires less than 200\,MiB.
For example, with a spline error of 16 (instead of 2) and 20 instead of 25 radix bits, RS trades performance (-11.5\%) for a significant space reduction (-99.9\%).

\begin{figure}
\centering
\includegraphics[width=0.75\linewidth]{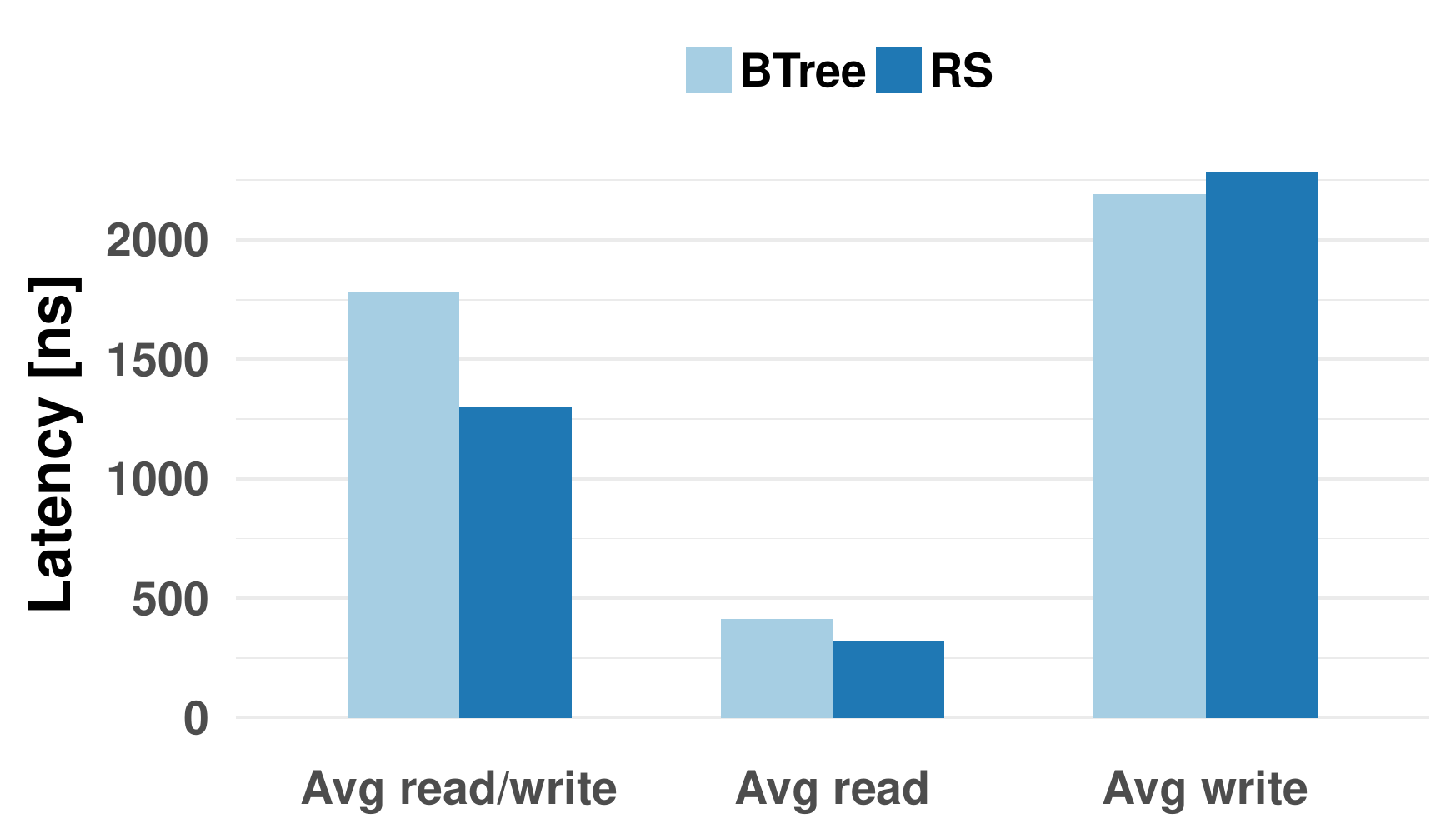}
\vspace{-4mm}
\caption{Average operator cost, LSM-tree.}
\vspace{-6mm}
\label{fig:lsm}
\end{figure}

{\ParHead LSM Performance.} To validate the applicability of RS to LSMs, we performed a preliminary experiment where we substitute the BTree index with a RadixSpline in RocksDB. We use the \texttt{osmc} dataset and executed 400\,M operations, 50\% reads and 50\% writes (cf. Figure~\ref{fig:lsm}). When using RS, the average write time increased by $\approx 4\%$, but the average read time decreased by over 20\%. The total execution time fell to 521 seconds from 712 seconds with a BTree. In addition, the RS variant used $\approx 45\%$ less memory, potentially creating space for larger Bloom filters or increased caching. While obviously preliminary, this experiment indicates potential benefits of RS in LSMs.

%% file: conclusions.tex
\section{Conclusions}
\label{sec:conclusion}

We have described a new learned index, called RadixSpline, that can be built in a single pass over sorted data.
Notably, RS only takes two hyper parameters and thus is rather easy to tune to a given dataset and memory budget.
Our experiments with real-world data have shown that RS is competitive with a state-of-the-art learned index in size and lookup performance while being as efficient to build as traditional indexes.
While our radix table has a constant size, it may become less useful as dataset size grows, or under large outliers.

In future work, we also plan to investigate how RS can be automatically tuned with minimal user-interaction, balancing memory footprint and performance.
Such an auto-tuning could be informed by metrics extracted from the data.
RS currently does not take advantage of any multi-threading, another potential direction for performance improvements.
\clearpage

{\ParHead Acknowledgments.}
This research is supported by Google, Intel, and Microsoft as part of the MIT Data Systems and AI Lab (DSAIL) at MIT, NSF IIS 1900933, DARPA Award 16-43-D3M-FP040, and the MIT Air Force Artificial Intelligence Innovation Accelerator (AIIA).

\balance

%% file: main.bbl
\begin{thebibliography}{10}

\bibitem{url-stxbtree}
{{STX B}}+ {{Tree}}, {{https://panthema.net/2007/stx-btree/}}.

\bibitem{hot}
R.~Binna, E.~Zangerle, M.~Pichl, G.~Specht, and V.~Leis.
\newblock {{HOT}}: {{A}} height optimized trie index for main-memory database
  systems.
\newblock In {\em Proceedings of the 2018 International Conference on
  Management of Data}, {{SIGMOD}} '18, pages 521--534, {New York, NY, USA},
  2018. {Association for Computing Machinery}.

\bibitem{alex}
J.~Ding, U.~F. Minhas, H.~Zhang, Y.~Li, C.~Wang, B.~Chandramouli, J.~Gehrke,
  D.~Kossmann, and D.~Lomet.
\newblock {{ALEX}}: {{An Updatable Adaptive Learned Index}}.
\newblock {\em arXiv:1905.08898 [cs]}, May 2019.

\bibitem{start}
P.~{Fent}, M.~{Jungmair}, A.~{Kipf}, and T.~{Neumann}.
\newblock {START — Self-Tuning Adaptive Radix Tree}.
\newblock In {\em 2020 IEEE 36th International Conference on Data Engineering
  Workshops (ICDEW)}, pages 147--153, 2020.

\bibitem{pgm-index}
P.~Ferragina and G.~Vinciguerra.
\newblock The {{PGM}}-index: A fully-dynamic compressed learned index with
  provable worst-case bounds.
\newblock {\em Proceedings of the VLDB Endowment}, 13(8):1162--1175, Apr. 2020.

\bibitem{fiting_tree}
A.~Galakatos, M.~Markovitch, C.~Binnig, R.~Fonseca, and T.~Kraska.
\newblock {{FITing}}-{{Tree}}: {{A Data}}-aware {{Index Structure}}.
\newblock In {\em Proceedings of the 2019 {{International Conference}} on
  {{Management}} of {{Data}}}, {{SIGMOD}} '19, pages 1189--1206, {New York, NY,
  USA}, 2019. {ACM}.

\bibitem{pillars}
J.~Gottschlich, A.~{Solar-Lezama}, N.~Tatbul, M.~Carbin, M.~Rinard,
  R.~Barzilay, S.~Amarasinghe, J.~B. Tenenbaum, and T.~Mattson.
\newblock The three pillars of machine programming.
\newblock In {\em Proceedings of the 2nd {{ACM SIGPLAN International Workshop}}
  on {{Machine Learning}} and {{Programming Languages}}}, {{MAPL}} 2018, pages
  69--80, {Philadelphia, PA, USA}, June 2018. {Association for Computing
  Machinery}.

\bibitem{ibtree}
G.~Graefe.
\newblock B-tree indexes, interpolation search, and skew.
\newblock In {\em Proceedings of the 2nd International Workshop on {{Data}}
  Management on New Hardware}, {{DaMoN}} '06, {Chicago, Illinois}, June 2006.
  {Association for Computing Machinery}.

\bibitem{fast}
C.~Kim, J.~Chhugani, N.~Satish, E.~Sedlar, A.~D. Nguyen, T.~Kaldewey, V.~W.
  Lee, S.~A. Brandt, and P.~Dubey.
\newblock {{FAST}}: Fast architecture sensitive tree search on modern {{CPUs}}
  and {{GPUs}}.
\newblock In {\em Proceedings of the 2010 {{International Conference}} on
  {{Management}} of {{Data}}}, {{SIGMOD}} '10, 2010.

\bibitem{deep_card_est2}
A.~Kipf, T.~Kipf, B.~Radke, V.~Leis, P.~Boncz, and A.~Kemper.
\newblock Learned {{Cardinalities}}: {{Estimating Correlated Joins}} with
  {{Deep Learning}}.
\newblock In {\em 9th {{Biennial Conference}} on {{Innovative Data Systems
  Research}}}, {{CIDR}} '19, 2019.

\bibitem{sosd}
A.~Kipf, R.~Marcus, A.~{van Renen}, M.~Stoian, A.~Kemper, T.~Kraska, and
  T.~Neumann.
\newblock {{SOSD}}: {{A Benchmark}} for {{Learned Indexes}}.
\newblock In {\em {{ML}} for {{Systems}} at {{NeurIPS}}}, {{MLForSystems}} @
  {{NeurIPS}} '19, Dec. 2019.

\bibitem{deep_sketch}
A.~Kipf, D.~Vorona, J.~M{\"u}ller, T.~Kipf, B.~Radke, V.~Leis, P.~Boncz,
  T.~Neumann, and A.~Kemper.
\newblock Estimating {{Cardinalities}} with {{Deep Sketches}}.
\newblock In {\em Proceedings of the 2019 {{International Conference}} on
  {{Management}} of {{Data}}}, {{SIGMOD}} '19, pages 1937--1940, {Amsterdam,
  Netherlands}, June 2019. {Association for Computing Machinery}.

\bibitem{ml_index}
T.~Kraska, A.~Beutel, E.~H. Chi, J.~Dean, and N.~Polyzotis.
\newblock The {{Case}} for {{Learned Index Structures}}.
\newblock In {\em Proceedings of the 2018 {{International Conference}} on
  {{Management}} of {{Data}}}, {{SIGMOD}} '18, pages 489--504, {New York, NY,
  USA}, 2018. {ACM}.

\bibitem{art}
V.~Leis, A.~Kemper, and T.~Neumann.
\newblock The adaptive radix tree: {{ARTful}} indexing for main-memory
  databases.
\newblock In {\em Proceedings of the 2013 {{IEEE}} International Conference on
  Data Engineering}, {{ICDE}} '13, pages 38--49, {USA}, 2013. {IEEE Computer
  Society}.

\bibitem{lsm_survey}
C.~Luo and M.~J. Carey.
\newblock {{LSM}}-based storage techniques: A survey.
\newblock {\em PVLDB}, 29(1):393--418, Jan. 2020.

\bibitem{bao}
R.~Marcus, P.~Negi, H.~Mao, N.~Tatbul, M.~Alizadeh, and T.~Kraska.
\newblock Bao: {{Learning}} to {{Steer Query Optimizers}}.
\newblock {\em arXiv:2004.03814 [cs]}, Apr. 2020.

\bibitem{neo}
R.~Marcus, P.~Negi, H.~Mao, C.~Zhang, M.~Alizadeh, T.~Kraska, O.~Papaemmanouil,
  and N.~Tatbul.
\newblock Neo: {{A Learned Query Optimizer}}.
\newblock {\em PVLDB}, 12(11):1705--1718, 2019.

\bibitem{rejoin}
R.~Marcus and O.~Papaemmanouil.
\newblock Deep {{Reinforcement Learning}} for {{Join Order Enumeration}}.
\newblock In {\em First {{International Workshop}} on {{Exploiting Artificial
  Intelligence Techniques}} for {{Data Management}}}, {{aiDM}}@{{SIGMOD}} '18,
  {Houston, TX}, 2018.

\bibitem{cidr_dlqo}
R.~Marcus and O.~Papaemmanouil.
\newblock Towards a {{Hands}}-{{Free Query Optimizer}} through {{Deep
  Learning}}.
\newblock In {\em 9th {{Biennial Conference}} on {{Innovative Data Systems
  Research}}}, {{CIDR}} '19, 2019.

\bibitem{cdfshop}
R.~Marcus, E.~Zhang, and T.~Kraska.
\newblock {{CDFShop}}: {{Exploring}} and {{Optimizing Learned Index
  Structures}}.
\newblock In {\em Proceedings of the 2020 {{ACM SIGMOD International
  Conference}} on {{Management}} of {{Data}}}, {{SIGMOD}} '20, {Portland, OR},
  June 2020.

\bibitem{plan_loss}
P.~Negi, R.~Marcus, H.~Mao, N.~Tatbul, T.~Kraska, and M.~Alizadeh.
\newblock Cost-{{Guided Cardinality Estimation}}: {{Focus Where}} it
  {{Matters}}.
\newblock In {\em Workshop on {{Self}}-{{Managing Databases}}}, {{SMDB}} @
  {{ICDE}} '20, 2020.

\bibitem{spline}
T.~Neumann and S.~Michel.
\newblock Smooth interpolating histograms with error guarantees.
\newblock In {\em Sharing Data, Information and Knowledge, 25th {{British
  National Conference}} on {{Databases}}}, {{BNCOD}} '08, pages 126--138, 2008.

\bibitem{lsm}
P.~O'Neil, E.~Cheng, D.~Gawlick, and E.~O'Neil.
\newblock The log-structured merge-tree ({{LSM}}-tree).
\newblock {\em Acta Informatica}, 33(4):351--385, June 1996.

\bibitem{qo_state_rep}
J.~Ortiz, M.~Balazinska, J.~Gehrke, and S.~S. Keerthi.
\newblock Learning {{State Representations}} for {{Query Optimization}} with
  {{Deep Reinforcement Learning}}.
\newblock In {\em 2nd {{Workshop}} on {{Data Managmeent}} for {{End}}-to-{{End
  Machine Learning}}}, {{DEEM}} '18, 2018.

\bibitem{lis_func_interp}
N.~Setiawan, B.~Rubinstein, and R.~{Borovica-Gajic}.
\newblock Function {{Interpolation}} for {{Learned Index Structures}}.
\newblock In {\em Database {{Theory}} and {{Applications}}}, {{DTA}} '20, 2020.

\bibitem{sql_embed}
{Shrainik Jain}, {Jiaqi Yan}, {Thiery Cruanes}, and {Bill Howe}.
\newblock Database-{{Agnostic Workload Management}}.
\newblock In {\em 9th {{Biennial Conference}} on {{Innovative Data Systems
  Research}}}, {{CIDR}} '19, 2019.

\bibitem{learn_cost}
J.~Sun and G.~Li.
\newblock An end-to-end learning-based cost estimator.
\newblock {\em Proceedings of the VLDB Endowment}, 13(3):307--319, Nov. 2019.

\bibitem{skinnerdb}
I.~Trummer, S.~Moseley, D.~Maram, S.~Jo, and J.~Antonakakis.
\newblock {{SkinnerDB}}: {{Regret}}-bounded {{Query Evaluation}} via
  {{Reinforcement Learning}}.
\newblock {\em PVLDB}, 11(12):2074--2077, 2018.

\bibitem{ml_tuning}
D.~Van~Aken, A.~Pavlo, G.~J. Gordon, and B.~Zhang.
\newblock Automatic {{Database Management System Tuning Through Large}}-scale
  {{Machine Learning}}.
\newblock In {\em Proceedings of the 2017 {{ACM International Conference}} on
  {{Management}} of {{Data}}}, {{SIGMOD}} '17, pages 1009--1024, {New York, NY,
  USA}, 2017. {ACM}.

\bibitem{local_card_est}
L.~Woltmann, C.~Hartmann, M.~Thiele, D.~Habich, and W.~Lehner.
\newblock Cardinality estimation with local deep learning models.
\newblock In {\em Proceedings of the {{Second International Workshop}} on
  {{Exploiting Artificial Intelligence Techniques}} for {{Data Management}}},
  {{aiDM}} '19, pages 1--8, {Amsterdam, Netherlands}, July 2019. {Association
  for Computing Machinery}.

\bibitem{naru}
Z.~Yang, E.~Liang, A.~Kamsetty, C.~Wu, Y.~Duan, X.~Chen, P.~Abbeel, J.~M.
  Hellerstein, S.~Krishnan, and I.~Stoica.
\newblock Deep unsupervised cardinality estimation.
\newblock {\em Proceedings of the VLDB Endowment}, 13(3):279--292, Nov. 2019.

\bibitem{surf}
H.~Zhang, H.~Lim, V.~Leis, D.~G. Andersen, M.~Kaminsky, K.~Keeton, and
  A.~Pavlo.
\newblock {{SuRF}}: {{Practical Range Query Filtering}} with {{Fast Succinct
  Tries}}.
\newblock In {\em Proceedings of the 2018 {{International Conference}} on
  {{Management}} of {{Data}}}, {{SIGMOD}} '18, pages 323--336, {Houston, TX,
  USA}, May 2018. {Association for Computing Machinery}.

\end{thebibliography}
